\documentclass{article}

\usepackage{microtype}
\usepackage{graphicx}
\usepackage{subfigure}
\usepackage{amssymb}
\usepackage{amsmath}

\usepackage{todonotes}

\usepackage{textcomp}
\usepackage{pythonhighlight}

\usepackage{hyperref}

\usepackage{amsmath,amsfonts,bm}

\def\eqref#1{equation~\ref{#1}}

\def\1{\bm{1}}

\DeclareMathAlphabet{\mathsfit}{\encodingdefault}{\sfdefault}{m}{sl}
\SetMathAlphabet{\mathsfit}{bold}{\encodingdefault}{\sfdefault}{bx}{n}

\def\sB{{\mathbb{B}}}

\def\sR{{\mathbb{R}}}
\def\sS{{\mathbb{S}}}

\usepackage[accepted]{icml2020}

\icmltitlerunning{$\mathcal{T}_p\mathcal{G}$ Geoopt: Riemannian Optimization in PyTorch}

\begin{document}

\twocolumn[
\icmltitle{$\mathcal{T}_p\mathcal{G}$ Geoopt: Riemannian Optimization in PyTorch}




\begin{icmlauthorlist}
\icmlauthor{Max Kochurov}{sk}
\icmlauthor{Rasul Karimov}{sk}
\icmlauthor{Serge Kozlukov}{hse,sk}
\end{icmlauthorlist}

\icmlaffiliation{sk}{Skolkovo Institute of Science and Technology, Moscow, Russia}
\icmlaffiliation{hse}{HSE University, Moscow, Russia}

\icmlcorrespondingauthor{Max Kochurov}{maxim.v.kochurov@gmail.com}

\icmlkeywords{Machine Learning, ICML}

\vskip 0.3in
]



\printAffiliationsAndNotice{} 

\begin{abstract}
Geoopt is a research-oriented modular open-source package
for Riemannian Optimization in PyTorch.
The core of Geoopt is a standard \texttt{Manifold} interface
that allows for the generic implementation of optimization
algorithms~\cite{radam2018becigneul}.
Geoopt supports basic Riemannian SGD as well as adaptive optimization
algorithms. Geoopt also provides several algorithms and arithmetic methods
for supported manifolds, which allow composing geometry-aware
neural network~\cite{hnn2018ganea, hgnn2019liu, spdbatchnorm2019brooks} 
layers that can be integrated with existing models.
\end{abstract}

\section{Introduction}

Geoopt
is built on top of PyTorch~\cite{pytorch2019paszke}, a dynamic computation graph backend. This allows us to use all the capabilities of PyTorch for geometric deep learning, including auto-differentiation, GPU acceleration, and exporting models~(e.g., ONNX~\cite{onnx2019bai}). Geoopt optimizers implement the interface of native PyTorch optimizers and can serve as a drop-in replacement during training.
The only difference is how parameters are declared\footnote{More examples can be found here:  \url{https://github.com/geoopt/geoopt/tree/master/examples}}, see \autoref{snippet:params}. The created manifold parameters can be used transparently with PyTorch functions and its serialization utils.
All native PyTorch tensors by Geoopt optimizers are treated as regular Euclidean parameters.

\begin{figure}
\begin{python}
import geoopt
from geoopt.optim import (
    RiemannianAdam
)
manifold = geoopt.Stiefel()
orth_mat = geoopt.Parameter(
    manifold.random(10, 10)
)
opt = RiemannianAdam([orth_mat])
\end{python}
\caption{\label{snippet:params}Creation of a manifold valued parameter.}
\end{figure}

The work on the package is mostly motivated by experiments
with hyperbolic embeddings and hyperbolic neural networks.
We provide several models of hyperbolic space, including
the Poincar\`e ball model, the Hyperboloid model, and general \(\kappa\)-Stereographic model which
generalizes Hyperbolic, Euclidean, and Spherical geometries~\cite{constcur2019bachmann}. 


\section{Riemannian optimization}

\begin{figure}[h]\centering
\includegraphics[width=.9\linewidth]{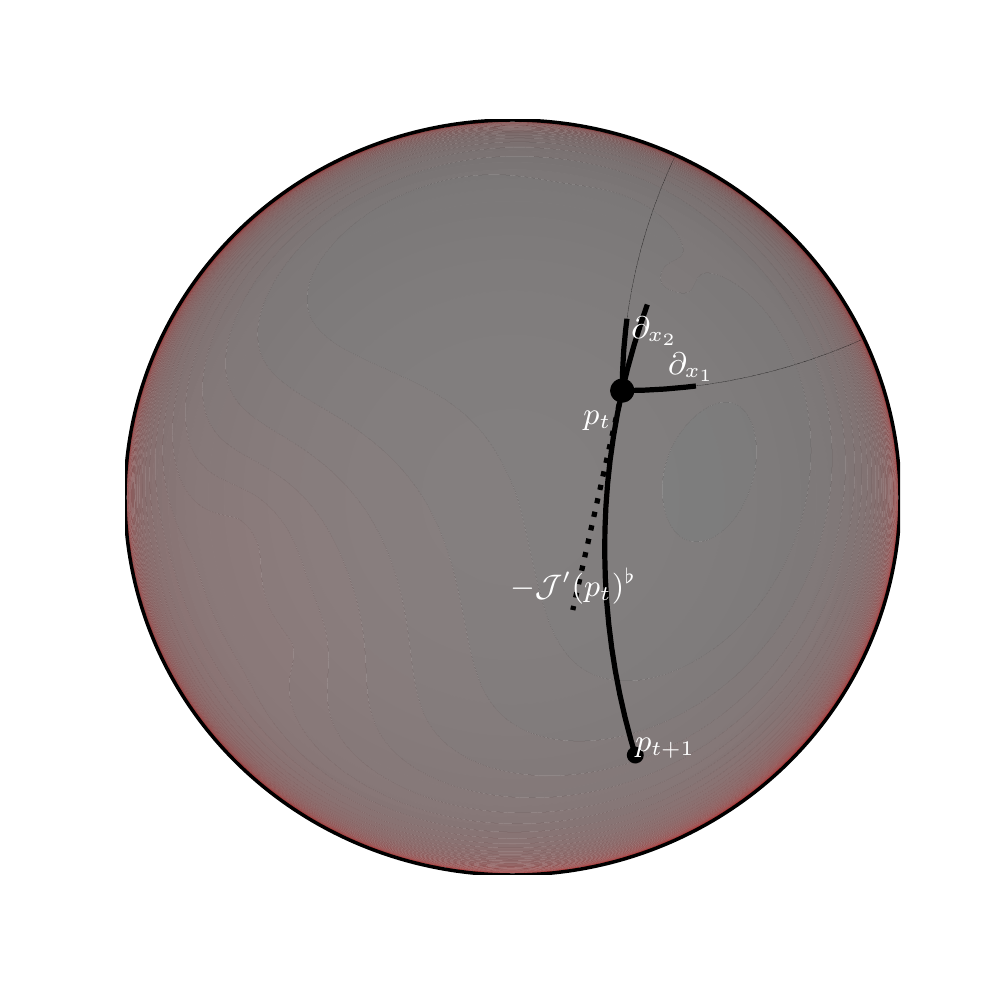}
\caption{A gradient descent step on the Poincar\'e disk.
Contour lines visualize the objective function; \( p_t \) is the current estimate;
\( -\flat(\operatorname{d}\mathcal{J}) \) is the descent direction, visualized as a geodesic curve;
\( p_{t+1} \) is the final point of that curve and the new estimate;
\( \partial_{x_1},~\partial_{x_2} \) are basis vectors in the space of directions
at \( p_t \);
stroked line visualizes the (downscaled) ``Euclidean'' gradient.
}
\label{fig:rgdStep}
\end{figure}

For a thorough introduction to geometry and differential geometry
we refer to~\cite{gravityLight,geometricAnatomy,leeRiem,leeSmooth,thurstonThree},
for synthetic description in general metric spaces to~\cite{yokota2012rigidity},
and concerned specifically with optimization and automatic
differentiation~\cite{betanalphaDiffGeometry,matmanifolds2007absil,elliott2018simple,elliottbeautiful}.

\autoref{fig:rgdStep} visualizes a gradient descent step on the Poincar\'e disk.
The concept of ``directions'' on a manifold corresponds to length-minimizing paths emanating from a point.
Restricted to a single source point, these paths, in a delicate way,
form a vector space, denoted \( \mathcal{T}_pM \) and called the ``tangent space''
at point \( p \).
Given such a path segment \( X \), we can obtain its destination point using the operation
called ``exponential map'', \( p_{t+1} = \exp X \).
In a small neighbourhood, one can find a unique shortest path
connecting one point to another -- this is called the logarithmic map,
\( X = \log_{p_t} p_{t+1} \).
The linear approximation (the derivative) of a function
between manifolds is thus a linear map that takes directions
in the input manifold into directions on the output manifold.
For an objective function \( \mathcal{J}: M \to \mathbb{R} \)
this means that derivative at a point \( p_t \) is
an operator \( \mathcal{J}'(p_t): \mathcal{T}_{p_t} M \to \mathbb{R} \),
i.e a linear functional. Given an inner product (a Riemannian local metric)
\( g_{p_t} \), there is unique direction \( \mathcal{J}'(p_t)^\flat \in \mathcal{T}_{p_t}M \)
that corresponds to this linear functional, in such a way that
\( \mathcal{J}'(p_t) = g_{p_t}\left(\mathcal{J}'(p_t)^\flat,~{\cdot}\:\right) \), assuming convenient placeholder notation.
It is the sought for ascent direction.
Thus the update rule is
\[ p_{t+1} = \exp(-\eta \mathcal{J}'(p_t)^\flat),\]
where \( \eta\in\mathbb{R} \) is the learning rate.

In Geoopt, points and directions are numerically represented
using embeddings of manifolds into ambient vector spaces
(often embedding is the identity map).
Objective functions, too,
are defined in this ambient space. Using PyTorch's
\texttt{backward} we can obtain the derivative of this
``extended'' function, acting on ``Euclidean'' directions.
As an embedding map allows to ``push'' a direction on the manifold
into a direction in the ambient vector space, this ``Euclidean''
derivative naturally corresponds to a linear functional
acting on directions on the manifold (which pushes
directions to ambient space and applies the ``Euclidean'' derivative).
This functional is exactly the derivative of our original objective function
defined on the manifold, and we can use the inner product
to convert it into the ascent direction, as discussed earlier.
This whole procedure -- the transition from ambient space
to the manifold, and application of inner product -- is performed
in Geoopt by a single operation, \texttt{egrad2rgrad}.

\section{Design goals}

Optimization on manifolds is a fairly general problem and designing
a general-purpose package accounting for possible use-cases may
not be a tractable problem. Geoopt is specifically concerned with
geometric deep learning \emph{research}
and its development is guided by a couple of rather pragmatic principles:

\begin{enumerate}
    \item \textbf{Smooth integration with the PyTorch ecosystem.}
    This assumes ``familiar'' PyTorch-esque interfaces.
    For instance, \texttt{geoopt.optim} optimizers can serve
    as drop-in replacements of \texttt{torch.optim}.
    This also implies compatibility with third-party
    packages based on PyTorch, for example, experiment management systems~\cite{falcon2019pytorch,catalyst}.
    \item \textbf{Broadcasting.} Support broadcasting for all operations and broadcasting semantics for product manifolds.
    \item \textbf{Robustness and numerical stability.} Hyperbolic models such as Poincaré disk and the Lorentz model have an unbounded numerical error as points get far from the origin. Therefore it is important that Geoopt users don't have to deal with more \texttt{NaN}s that they would have to otherwise. Whenever possible, algorithms in Geoopt are implemented to work even with \texttt{float32} precision. The instabilities of specific functions are described in documentation appropriately.
    \item \textbf{Efficiency and extendibility.} The previous bullets are concerned with ``not getting in the way''.
    When those are satisfied, we strive to provide reasonable efficiency and leave place for extendibility.
\end{enumerate}

\section{Implementation details}

The basic primitive of Geoopt is \texttt{geoopt.ManifoldTensor}
which is a ``tensor'' (a multi-dimensional array)
that stores a reference to its containing \texttt{geoopt.Manifold}.
We inherit from \texttt{torch.Tensor} and \texttt{torch.nn.Parameter}.
This ensures compatibility with the rest of PyTorch ecosystem
and suggests just one ``right way'' to use Geoopt within PyTorch code,
which we consider Pythonic~\cite{pep8}.

Array manipulations in Geoopt should support broadcasting.
Simple product manifolds are implemented with broadcast along first dimensions, by convention.
More complex cases are handled by \texttt{geoopt.ProductManifold} class.

The original goal of Geoopt is Riemannian optimization, and it is supposed to be efficient:
this requires optimizations in the update step,
merging retractions followed by parallel transport, etc.
In product manifolds, the adaptive term is computed per manifold parameter, and product structure is exploited~\cite{radam2018becigneul}.
This is a part of Geoopt in the first place, and any possibility to make effective use of the adaptive term is implemented. 

The \texttt{geoopt.Manifold} base class describes a methodset expected
by \texttt{geoopt.optim} optimizers. The \texttt{geoopt.Manifold} inherits
from \texttt{torch.nn.Module}: this way it is captured by \texttt{state\_dict()}
and its parameters can be optimized for.

The minimal methodset for the \texttt{geoopt.Manifold} subclass includes:
\begin{itemize}
\item \texttt{Retr}action: \( \operatorname{retr} \)
takes an array of points, an array of tangent vectors
at these points, and outputs an array of points.
Retraction is a first-order approximation of the exponential map
used in optimization, and often we have a separate \texttt{expmap} method.
However, for some manifolds, we provide variants that perform
the actual exponential map instead of retraction during optimization.
\item Vector \texttt{transp}ort: \( \operatorname{transp} \)
takes an array of source points, an array of target points,
an array of tangent vectors attached to source points,
and produces an array of tangent vectors at target points.
It is the first-order approximation of parallel transport.
\item \texttt{Inner} product:
\( \operatorname{inner} \) takes an array of points
and two arrays of tangent vectors at these points
and returns an array of inner products of those vectors.
\item \texttt{egrad2rgrad} is used to convert
the covector in the ambient vector space (produced by PyTorch's \texttt{backward})
into a corresponding tangent vector on the actual manifold.
\end{itemize}

Points and tangent vectors in Geoopt are always represented by coordinates in the (assumed) ambient vector space.
In case of \texttt{PoincareBall},
the embedding coincides with the natural global chart,
and corresponds to the chart-induced basis vector fields.
Such consistency is only possible because of
negative curvature of Hyperbolic space and conformality of Poincar\'e Ball.
On a sphere, one could neither allocate a non-vanishing smooth
vector field, nor expect unique geodesics to exist between all points, nor measures to
have unique barycentres. For this reason, on a Sphere
one has to either use local charts
or take on the extrinsic approach (assume an ambient vector space, which is what we do).
The array of numbers representing a tangent vector
(e.g., one gets after taking a logarithmic map) in Geoopt stores
the coordinates of the push-forward of that vector under the assumed
embedding into ambient vector space.
This representation is somewhat restrictive (e.g., it complicates
implementing the tiling-based parameterizations of Hyperbolic
space~\cite{yuSaTilingBased}) but rather convenient
and follows the spirit of~\cite{radam2018becigneul}.

To extend Geoopt, one should implement basic methods such as retraction or exponential map
on the manifold, parallel or vector transport for tangent vectors, and make them properly broadcastable.
The latter might be the hardest in implementation, and as maintainers, we are more than ready to help with it.
\section{Features}
To help researches Geoopt has implementation of standard manifolds~\cite{matmanifolds2007absil}:
\begin{itemize}
    \item \texttt{geoopt.Sphere} manifold -- for unit norm constrained problems (embeddings, eigenvalue problems)
    \begin{equation}
        \sS = \left\{x \in \sR^n: \|x\| = 1\right\}
    \end{equation}
    \item \texttt{geoopt.Stiefel} manifold -- for basis reconstruction
    \begin{equation}
        \mathrm{St} = \left\{X \in \sR^{n\times m} :  X^\top X = I \right\}
    \end{equation}
    \item \texttt{geoopt.BirkhoffPolytope}~\cite{Douik2018Manifold} -- for inferring permutations in data
    \begin{equation}
        \sB = \left\{X \in \sR^{n\times n}:  \mathbf{1}^\top X = \mathbf{1} = X\mathbf{1}\right\}
    \end{equation}
    \item \texttt{geoopt.Stereographic} model~\cite{constcur2019bachmann} and \texttt{geoopt.Lorentz} manifold -- for Hyperbolic deep learning
    \item \texttt{geoopt.Product} and \texttt{geoopt.Scaled} manifolds -- to combine and extend any of above
\end{itemize}
Geoopt supports most important and widely used optimizers:
\begin{itemize}
    \item \texttt{geoopt.optim.RiemannianAdam} -- a Riemannian version for popular Adam optimizer \cite{adam2014kingma}
    \item \texttt{geoopt.optim.SparseRiemannianAdam} -- Adam implementation to support sparse gradients
    \item \texttt{geoopt.optim.RiemannianSGD} -- SGD with (Nesterov) momentum implementation
    \item \texttt{geoopt.optim.SparseRiemannianSGD} -- SGD implementation that supports sparse gradients
\end{itemize}
\section{Advanced Usage}
The advanced usage of Geoopt covers Hyperbolic deep learning pioneered in recent years \cite{spacetimelocal2015sun,fairpoincare2017,representation2018desa,hypgroups87,embeddtext2018dhingra}. 
In Geoopt, we provide a robust implementation for the Poincare Ball model along with methods for performing supplementary math.
In addition to constant negative curvature support, positive curvature stereographic model of a sphere is also a part of the unified implementation of M\"obius arithmetics in projected spacetime domain. Users can find supplementary functions as methods of \texttt{geoopt.Stereographic} class. Derivatives for curvature are supported by the whole domain, especially for zero curvature case, so curvature optimization is possible.

\subsection{Other Applications}

Geoopt is a general-purpose optimization library for PyTorch. Manifold optimization appears in many applications. 

\paragraph{Language models.} For example, in NLP, when training recurrent neural networks, it is useful to constraint the transition matrix to be unitary~\cite{arjovsky2015unitary}. The unitary matrix keeps the gradient norm unchanged, and the network is able to learn long-range dependencies. Unitary matrices form a smooth Riemannian manifold, and Riemannian optimization can be easily applied to them. Another kind of constrained parameterization used in RNNs is Stiefel manifold~\cite{helfrich2017orthogonal}. It also helps to avoid problems of vanishing or exploding gradients.

\paragraph{Computer vision.} In the field of computer vision, doubly stochastic matrices can be used to match keypoints between views~\cite{birdal2019probabilistic}. In~\cite{birdal2019probabilistic} the probabilistic approach was proposed to compare images from a completely different time and viewpoints. To calculate uncertainty bounds, MCMC is run over the solution space. Combined with cycle consistency energy function method is available not only to match keypoints but also to provide estimates guiding to pick the most promising connections.

\paragraph{Time series.} For multidimensional time series analysis and classification, it was shown promising to look at the covariance matrix of stationary representation. The covariance matrix is passed to SPD neural networks that perform final classification~\cite{gesturespd2019xuan,spdbatchnorm2019brooks}, e.g., processes or gestures. The approach proposed in~\cite{spdbatchnorm2019brooks} allows Riemannian batch normalization for SPD matrices, further improving time series classification benchmarks and training stability.

\paragraph{Hyperbolic deep learning.}
An active area of research is using hyperbolic representations to account for ``implicit hierarchical relationships'' in data.
Geoopt allows for optimization with parameters in several models of real Hyperbolic spaces,
and provides basic operations of hyperbolic geometry.
Hyperbolic embeddings appear in NLP~\cite{mrelpoincare2019balazevic,fairpoincare2017}, image understanding~\cite{hyperbolic2019khrulkov}, and general representation learning~\cite{mixedcur2019paszke}.
Some works also focus on graph learning tasks~\cite{hgcn2019chami,hgnn2019liu,constcur2019bachmann}
and extend the message passing framework proposed by~\cite{torch_geometric2019fey}.
With Geoopt, implementation of such extensions become simpler, as demonstrated by~\cite{hgcn2019chami}.
An extensible implementation of Hyperbolic message passing framework may rely on \texttt{torch\_geometric} library modifying \texttt{aggregate} method in \texttt{MessagePassing} class.

\paragraph{Summary.}
Riemannian optimization is important for current research in geometric deep learning. Geoopt tries to fill the niche of Riemannian optimization in PyTorch.
The library has helped to conduct research in computer vision~\cite{hyperbolic2019khrulkov,birdal2019probabilistic,chen2019hyperbolic}, navigation~\cite{sar2020comer}, optimal transport~\cite{slicedgromovwasserstein2019vayer}, time-series analysis~\cite{time2020vayer}, and Hyperbolic deep learning~\cite{actions2020shen,skopek2020mixedcurvature,othierarchy2020melis,hgcn2019chami}.

\section{Related projects}

There were other Riemannian optimization projects prior to Geoopt.
Notable examples include PyManOpt~\cite{pymanopt} and GeomStats~\cite{geomstats}.
The main distinction between Geoopt and other solutions is interface-wise.
PyManOpt is a Python re-implementation of the original Manopt~\cite{manopt}
and follows the original interface closely
with its \texttt{solver.solve(Problem(manifold, cost))} semantics.
PyManOpt currently provides an admittedly broader collection of algorithms
(trusted region methods, Nelder-Mead, etc) and manifolds than Geoopt.
Manopt is the MATLAB package accompanying the Absil's book~\cite{matmanifolds2007absil}.
Geomstats is designed around sklearn's \texttt{fit-transform} semantics.
Both solutions are great general-purpose tools for Riemannian optimization.
Geoopt is concerned explicitly with neural networks and  geometric deep learning:
its interfaces are designed to integrate well with PyTorch-based projects.
Geoopt users define neural networks and cost functions in the usual ``PyTorch''
way and don't have to construct a PyManOpt \texttt{Problem}.
In this aspect, similar to Geoopt is McTorch~\cite{meghwanshi2018mctorch}.
It takes on the approach of forking PyTorch and extending it on the C++ back-end side.
This is heavy on infrastructure.
Maintaining a fork up to date demands a considerable and continuous effort.
Using a fork complicates integration with other third-party libraries,
which could pin to specific versions of PyTorch.
It could complicate it to the point that one runs into the task
of re-compilation of entire PyTorch and further distribution of binary packages.
Geoopt avoids such infra-structural costs and aims to keep the bar low
-- both for new contributors and users.


\bibliography{main.bib}
\bibliographystyle{icml2020}

\end{document}